# Trapped Ion Chain Thermometry and Mass Spectrometry through Imaging

**V Rajagopal‡, J P Marler §, M G Kokish, B C Odom**
Northwestern University, Evanston, IL
Email: b-odom@northwestern.edu

**Abstract.** We demonstrate a spatial-imaging thermometry technique for ions in a one-dimensional Coulomb crystal by relating their imaged spatial extent along the linear radiofrequency ion trap axis to normal modes of vibration of coupled oscillators in a harmonic potential. We also use the thermal spatial spread of 'bright' ions in the case of a two-species mixed chain to measure the center-of-mass (COM) resonance frequency of the entire chain and infer the molecular composition of the co-trapped 'dark' ions. These non-destructive techniques create new possibilities for better understanding of sympathetic cooling in mixed-ion chains, improving few-ion mass spectrometry, and trapped-ion thermometry without requiring a scan of Doppler cooling parameters.

PACS numbers: 37.10.Ty, 37.20.+j

*Keywords*: Ion trap, Thermometry, Mass spectrometry, Fluorescence imaging, Driven harmonic oscillator, Brownian motion

## 1. Introduction

Trapped laser-cooled atomic ions and co-trapped atomic or molecular species undergo a phase transition from a liquid to a Wigner crystal when their thermal energy becomes sufficiently small compared with their Coulomb interaction energy [1]. The fluorescence images of these atomic ions can be used to obtain a wealth of information ranging from the trap characteristics [2, 3] and temperature of the constituent ions [4], to the molecular composition [5, 6, 7] and internal state of the co-trapped ions [8]. For ions trapped in a harmonic potential undergoing thermal Brownian motion, the imaged spatial fluorescence distribution follows a Gaussian function [9, 10, 11]. In addition to the point-spread width from the imaging optics [12], the width of this Gaussian function contains contributions from the finite temperature of the ions [11, 13]. Using high magnification optics (~100), measurement of the time-averaged spatial distributions has previously been used to make the most precise temperature measurements of a single ion in the weak-binding regime [11]. In contrast to the more traditional method of measuring the fluorescence line shape, the spatial-imaging thermometry approach does not require separation of Lorentzian lifetime and thermal contributions to the line shape, where understanding the latter requires assumptions to be made about the thermal distribution [4]. Here, we extend this spatial-imaging thermometry technique for determination of the temperature of multiple ions in a linear Coulomb crystal, by relating their imaged spatial spreads to axial normal modes of a one-dimensional system of coupled oscillators.

‡ Present address: The Rowland Institute at Harvard, Cambridge, Massachusetts
§ Present address: Clemson University, Clemson, South Carolina

We also use imaging of the time-averaged spatial distributions to determine the mass of non-fluorescing species in a two-species crystal, with m/δm precision of 200. The apparatus for converting some fluorescing atomic ions into non-fluorescing molecular ions, described elsewhere [14, 15], allows us to concentrate the flow of reactant gas at the site of the Coulomb crystal while keeping the overall pressure in the chamber low. In order to determine the mass of the trapped molecular products of the chemical reaction, we use a count of the remaining fluorescing ion number, along with a measurement of the chain's center of mass axial frequency. The COM frequency is measured by monitoring the response of the imaged ion spatial widths to a swept-frequency rf drive applied to the trap electrodes.

**Apparatus**

Our experimental setup housed in a UHV chamber consists of a linear radiofrequency ion trap, with post-style endcaps as described in [16]. Axial confinement is provided by endcaps with $2z_0$ = 17.8 mm, and radial confinement is from a radiofrequency drive voltage applied to rod electrodes with an inscribed radius $2r_0$ = 9.2 mm and a $r_e/r_0$ = 0.89; $r_e$ is the radius of the rod electrodes. This design of $r_e/r_0$ allows a larger solid angle for light collection compared to conventional geometry [17], without compromising the trapping efficiency. A neutral barium beam produced by an oven, is ionized into $Ba^+$ in the trap using one of three two-photon schemes [18, 19, 20]. A $Ba^+$ Doppler cooling laser at 493 nm, red-detuned from the $|6S_{1/2}\rangle \leftrightarrow |6P_{1/2}\rangle$ transition, is split into two beams, with one propagating along the trap axis the other propagating at 45 ° relative to the two radial principle axes. Additionally a repump laser at 650 nm, on resonance with the $|5D_{3/2}\rangle \leftrightarrow |6P_{1/2}\rangle$ transition, propagates along the axis of the trap and overlaps with the 493 nm beam. A narrow collimated pulsed gas beam characterized in [15] chemically reacts with the $Ba^+$ atomic ions to convert some fraction to molecular ions.

Light from the fluorescing $Ba^+$ ions is focused with a magnification of ~10 onto an EMCCD camera with 512 × 512 pixels, yielding an image as shown in Fig. 1. The EMCCD camera is cooled to -70 degrees C, operated with a gain of 200, and uses an exposure time set to 800 ms. The trap electrode geometry provides optical access of ±20 degrees along the more tightly restricted radial direction, with light collected by optics (NA = 0.45) placed 7 cm from the trap center.

The imaging magnification is determined by comparing the pixel distance between ion centers in a 1D chain (see Fig. 2) to the theoretical center positions calculated in [21], given an axial COM frequency. With sufficient averaging (using 500 images taken with 800 ms exposure times) ion center positions were measured in pixels to much better than 1%. However, since in our calibration data sets the applied endcap voltage was recorded rather than using a measured COM frequency, the possible modification of the COM frequency due to trapped charge on electrode surfaces (patch potentials) creates an uncertainty in the calibration. A tabulated set of measured COM frequencies for given endcap voltages, taken over the months for which data was collected, constrains this calibration error to be less than 1%. Our camera calibration result is then $1.67 \pm 0.02$ $\mu$m per EMCCD pixel. Uncertainty in the camera calibration is treated as a systematic uncertainty in temperature and point spread function measurements, discussed below.

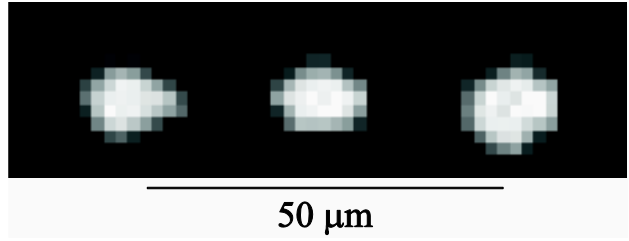

**Figure 1.** Typical single-shot EMCCD image of ion chain. Here, $\omega_{COM} = 2\pi \times 44$ kHz

## 2. Temperature measurement

The temperature at the Doppler cooling limit $T_D = \hbar\Gamma/(2k_B)$, where $\Gamma = 2\pi \times 15$ MHz is the natural linewidth of the transition, is 0.4 mK for Ba$^+$ cooled on the $|6S_{1/2}\rangle \leftrightarrow |6P_{1/2}\rangle$ transition. However, as is typical of experiments not attempting to cool to near the ground state of the trap, we expect ions in our short 1D chains to be at elevated temperatures of a few mK, due to non-optimized laser parameters and imperfect alignment of the ions along the field-free trap axis [23]. We operate in the weak binding limit, $\Gamma \gg \omega_z$, where $\omega_z/(2\pi)$ is the axial trapping frequency. In terms of the Doppler cooling limit, the weak binding limit corresponds to $k_B T_D/\hbar\omega_z \gg 1$, so many harmonic oscillator states are occupied. For a single ion in the weak binding limit, the spatial distribution is a Gaussian position distribution of width $\Delta z_{th}$ given by [10, 11]

$$f(z) = \sum_{n=0}^{\infty} P_n |\psi_n(z)|^2 = \frac{1}{\sqrt{2\pi}\Delta z_{th}} e^{\frac{-(z-z_0)^2}{2\Delta z_{th}^2}}, \quad (1)$$

where $z_0$ is the position of the ion center, n labels the harmonic oscillator quantum state, $P_n$ is the occupation probability, $\psi_n(z)$ is the normalized eigenfunction, and

$$\Delta z_{th} = \sqrt{\frac{k_B T}{m\omega_z^2}}. \quad (2)$$

We operate our cooling laser below saturation to avoid any distortion of the time-averaged line shape. The observed ion image is a broader Gaussian, also containing a contribution from the point spread function (PSF), due to the fact that a finite cone of light collected from a point source focuses onto a disc of finite radius [12]. The imaged spatial width of the ion $\Delta z$ is given by

$$\Delta z^2 = \Delta z_{th}^2 + \Delta z_{PSF}^2 \qquad (3)$$

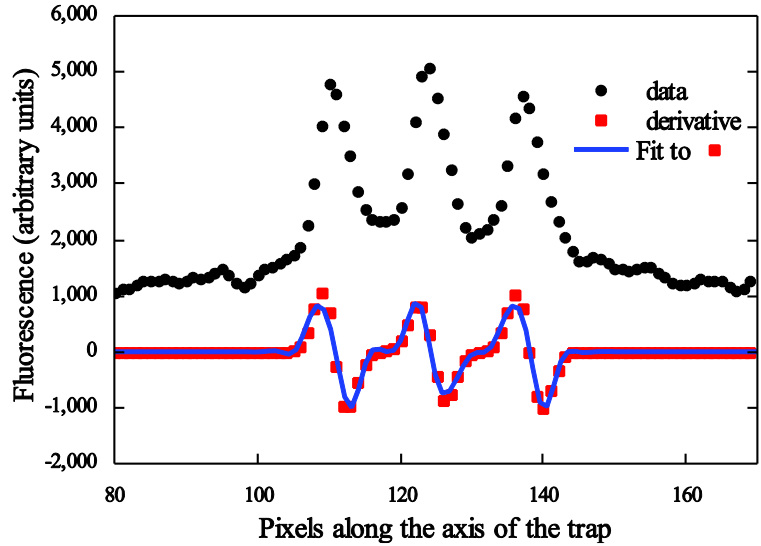

**Figure 2.** Fluorescence of a 3-ion chain at each position along the trap axis, summed within the region of interest over the transverse dimension (black dots). Here, $\omega_{COM} = 2\pi \times 39$ kHz. The spatial derivative of this fluorescence profile (red squares) is fit to a derivative of a multi-Gaussian function (blue line) to extract the spatial width and center position of each ion image.

For our operating conditions the axial confinement is intentionally made much weaker than the radial confinement, so the ions form a one-dimensional chain, and the radial and axial modes are to a good approximation decoupled [11, 21]. Our typical operating conditions are 1000 V peak-peak for the 3 MHz radiofrequency voltage and 20-60 V for the DC endcap voltage, resulting in $\omega_r \approx 2\pi \times 1$ MHz and $\omega_z = (2\pi) \times$ 20-45 kHz.

Trap imperfections such as uncompensated patch potentials or electrode misalignment can lead to micromotion excitation of the ions [22]. These effects are generally expected to be more noticeable for multiple ions, since potential compensation by tuning of shim electrodes cannot generally null micromotion for each ion in the sample. Besides the well-known micromotion heating of the secular modes, micromotion will also broaden the ion images along the direction of the rf field lines. However, since our analysis is performed on the imaged widths along the axial direction, for which rf fields are only due to small fringing fields, this broadening effect is expected to be very small for our short ion strings held at trap center. There is no evidence for this effect in our measurements.

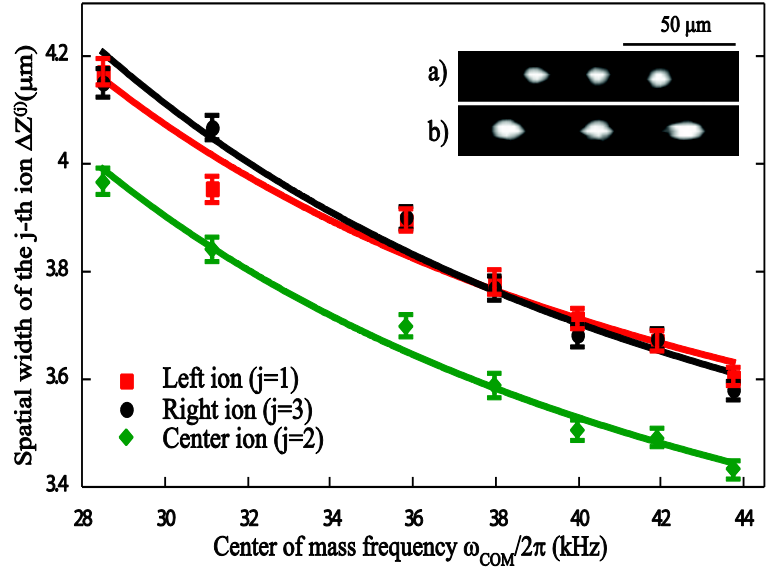

**Figure 3.** Imaged widths of ions in a three-ion chain versus axial COM frequency. The spatial width of the center ion is smaller than those of the end ions, as expected from Eq. 7. Each point represents the mean of a Gaussian fit to a distribution of 500 separate single-shot width measurements, with vertical bars representing the fit uncertainty. Every single-shot width measurement is obtained from a corresponding multi-Gaussian fit of ion fluorescence versus position, as shown in Fig. 2. Insets a) and b) are 73 × 11 pixel single-shot images of the three ions at high (~ 42 kHz) and low (~ 23 kHz) axial COM frequencies, respectively.

For J ions in a chain, the excursions of the j-th ion along the z-axis (axial direction) has contributions from J normal modes. The observed Gaussian width of the j-th ion is then

$$\Delta z^{(j)} = \sqrt{\left(\Delta z_{th}^{(j)}\right)^2 + (\Delta z_{PSF})^2}, \qquad (4)$$

where

$$\Delta z_{th}^{(j)} = \sqrt{s_1^{(j)} \times (k_B T/m\omega_1^2) + s_2^{(j)} \times (k_B T/m\omega_2^2) + \cdots} \qquad (5)$$

Note that all the axial normal modes are assumed to be in thermal equilibrium with one another at temperature T. Here, $\omega_p$ is the eigenfrequency of the p-th normal mode, and $S_p^{(j)}$ is the weighting factor for motion of the j-th ion resulting from excitation of the p-th mode, derived from the eigenvectors of normal modes, with $\sum_{p=1}^{J} S_p^{(j)} = 1$.

This equation can also be rewritten in terms of the lowest eigenfrequency of the system, $\omega_{COM}$, which for a single-species crystal precisely corresponds to center of mass oscillation in which all ions oscillate together with the same amplitude [21]. For a mixed- species crystal, correspondence between the COM frequency and a normal mode occurs in the strongly coupled limit, a condition we easily achieve by Doppler cooling the 1D chain and by using small RF excitation voltages, as discussed later. Even in the strongly coupled limit, the eigenmode of a multi-species crystal is slightly different from the COM mode due to configuration-dependent corrections [24], but the modifications to the eigenfrequency are well below our sensitivity. Neglecting the configuration-dependent corrections, the COM axial frequency of a chain is given by [7].

$$\frac{1}{\omega_{COM}^2} = \frac{\langle m \rangle}{\langle q \rangle} \frac{z_0^2}{2 \kappa U_{dc}} = \frac{\langle m \rangle}{q} \frac{z_0^2}{2 \kappa U_{dc}} \qquad (6)$$

where $U_{dc}$ is the voltage applied to the endcaps, and $\kappa$ is a geometric factor accounting for the fact that the electrode surfaces are not hyperbolas. Here $\langle m \rangle$ and $\langle q \rangle$ are the average mass and charge of ions in the chain, and the second equality is for the typical case where all the trapped ions have the same charge. Combining Eqs. 4 and 5, we express the ion spatial extent as

$$\Delta z^j = \sqrt{(\gamma^{(j)})^2 \ (k_B T / m \omega_{COM}^2) + (\Delta z_{PSF})^2} \qquad (7)$$

with $\gamma^{(j)}$ calculated according to [21, 14]. For a single ion, $\gamma^{(1)} = 1$, and in the case of a three-ion chain, the $\gamma^{(j)}$ values are 0.7271, 0.6695 and 0.7271 for the left, center, and right ion respectively. In general, the farther an ion is from the center of the chain, the larger its spatial spread. See Fig. 3 for observation of this relationship in a three-ion chain.

**Data Analysis**

In order to decrease the dead time between measurements, the data is binned transverse to the trap axis in the region of interest to create a 1D array of intensity values. Application of this same technique to characterization of radial widths of the ions would require separate binning of multiple segments of the image, one segment for each ion, and this quantitative analysis was not attempted. However, the expected qualitative behavior in the weak binding limit, that the radial widths do not appear to change even as the axial widths change with the axial confinement potential, can be observed even in single-shot images (see Fig. 3 insets).

Each 800 ms exposure yields a multi-peaked intensity distribution, with one peak per fluorescing ion, as shown in Fig. 2. The spatial width of each imaged ion is extracted from a fit to the derivative of this data. We find that fitting to the derivative tends to improve fit convergence in the presence of a nearly spatially constant scattering background and imperfect imaging. A multi-Gaussian function fits the data well, as expected [9, 10], and in practice we fit the derivative of the data to the derivative of a multi-Gaussian function.

For each ion in a string, a measurement of the imaged spatial extent versus $\omega_{COM}$ can be used in a 2-parameter fit to Eq. 7 to extract both a temperature and $\Delta z_{PSF}$. Fig. 3 shows such a curve for a 3-ion chain, with each data point given by an average of 500 multi-Gaussian fits to single-shot exposures.

Data were recorded from a three ion-crystal, corresponding to the image shown in Fig. 1, to perform three independent fits to measure T and $\Delta z_{PSF}$. The ion crystal temperatures determined from the spatial width versus COM frequency curves for the left, center and right ions are 7.2 mK, 8.4 mK, and 8.2 mK respectively, with corresponding statistical uncertainties of 0.7 mK, 0.5 mK and 0.7 mK. These statistical uncertainties are much larger than the 1% systematic uncertainty from camera calibration, and these independent temperature measurements are in good agreement with one another. The corresponding $\Delta z_{PSF}$ extracted from each of the three fits are 3.19 μm, 2.98 μm, and 3.10 μm. The fits are such that the PSF value is extracted from the data with higher precision than is temperature and the statistical uncertainty in the PSF is comparable to the 1% camera calibration systematic uncertainty. Adding these two uncertainties in quadrature yields 0.08 μm, 0.05 μm, and 0.08 μm for the three independent determinations of $\Delta z_{PSF}$. These extracted values for the PSF are in reasonable agreement with one another and are consistent with a PSF arising from imaging slightly (10%) worse than ideal diffraction-limited imaging for our optics.

In the three-ion case, the good agreement between the three independently measured temperatures helps to validate the spatial-imaging thermometry technique. We note that Fig. 3 shows quite different curves for the middle versus end ions, as they should be for thermal equilibrium, and that the extracted temperatures are in good agreement with one another. However, for chains longer than three to four ions, we observe a non-monotonic and non-reproducible relationship between the spatial width and the COM frequency, which becomes more significant for ions further from the center. We speculate that breakdown of the analysis technique is due to imperfect alignment, in the presence of patch potentials, of the chain with the trap axis and an associated mixing of high-frequency axial modes with radial modes. Further investigation would be required to investigate that hypothesis.

## 3. Mass Spectrometry

Secular excitation of two-species Coulomb crystals has previously been used to identify the mass of the non-fluorescing species (see, for instance [5, 6, 7] for representative examples). Here, rather than recording the fluorescence level of the entire sample, we determine the axial COM frequency by monitoring the spatial extent of individual fluorescing ions. Note that the axial COM frequency reflects the strength of the axial confinement potential and is quite distinct from the rf frequency applied to the rod electrodes.

Molecular ions produced in the trap as a result of barium ions reacting with a $N_2O$ jet are identified by the appearance of dark sites in the chain. Images of this chain are recorded as the frequency of a sinusoidal voltage applied to one of the endcaps is varied. For an applied drive frequency far away from the COM resonance of the chain, ions are well localized; closer to resonance, the ions are motionally excited and hence appear more spread out, as shown in the Fig. 4. The change in position of the bright ions from the on- versus off-resonance images is because of longer timescale hopping of the dark ions between sites.

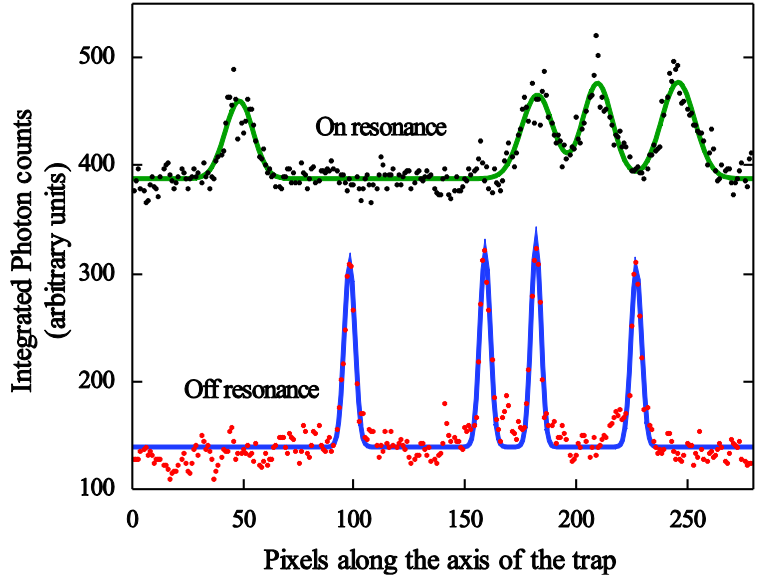

**Figure 4.** Ion spatial width response to an rf drive applied to an endcap. When excited by an rf voltage with frequency near the COM resonance, ions are less localized than when driven by an off-resonance drive. The on-and-off resonance images are taken with the same chain, but hopping reordered the crystal between the two sets of image captures. The two curves are offset vertically for clarity.

The average spatial spread of the bright laser-cooled ions is extracted as the drive frequency is varied using a multi-Gaussian fit. Strictly speaking, the spatial profile of each ion is a Gaussian only in the zero-drive limit and becomes symmetrically bimodal as the drive amplitude is increased beyond the thermal Brownian motion [25]. However, since we intentionally keep the drive strength weak, and furthermore we are only interested in extracting the central frequency of the resonance, a simple Gaussian fit for each ion suffices for our purposes in order to find the COM frequency. When plotted versus frequency, the COM response of the Coulomb crystal to a near-resonant drive should be well approximated by a Lorentzian line shape (Fig. 5). The molecular (dark) species is then identified from the shift in the resonant frequency as the composition of the crystal changes, as described below.

In the case of a strongly coupled two-species crystal [6] with $N_{\text{bright}}$ Ba$^+$ ions and $N_{\text{dark}}$ BaX$^+$ ions ($N_{\text{total}} = N_{\text{bright}} + N_{\text{dark}}$), Eq. 6 becomes

$$\frac{1}{\omega_{COM}^2} = \left(m_{Ba} + m_X \frac{N_{dark}}{N_{total}}\right) \frac{z_0^2}{2\kappa q U_{dc}} \tag{8}$$

where $N_{dark}/N_{total}$ is the fraction of dark ions in the crystal. Note that we have also assumed that the bright and dark species have the same charge; this condition is met for the case studied here, and is also quite often true for products of other reactions occurring between singly charged trapped ions and background gas. We resonantly excite the crystal after its composition is changed after each burst of gas, to get a series of response curves similar to Fig. 5. Averaging over 20 such curves after every burst, we obtain the data shown in Fig. 6. From the ratio of the slope to the intercept, the mass of the atom "X" in the molecular ion "BaX$^+$" is identified. For the case of a Ba$^+$/BaO$^+$ crystal, this technique leads to a 0.5% mass determination of the molecular mass, or a m/$\delta$m = 200.

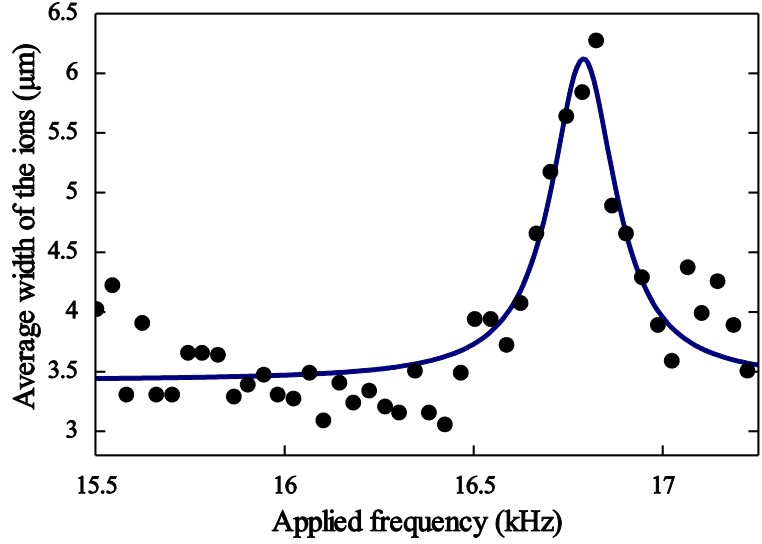

**Figure 5.** Response of the average ion spatial width to the swept-frequency rf drive applied to the endcaps. The spatial spread of the ions increases closer to the COM resonance frequency. A Lorentzian fit to the data is also plotted.

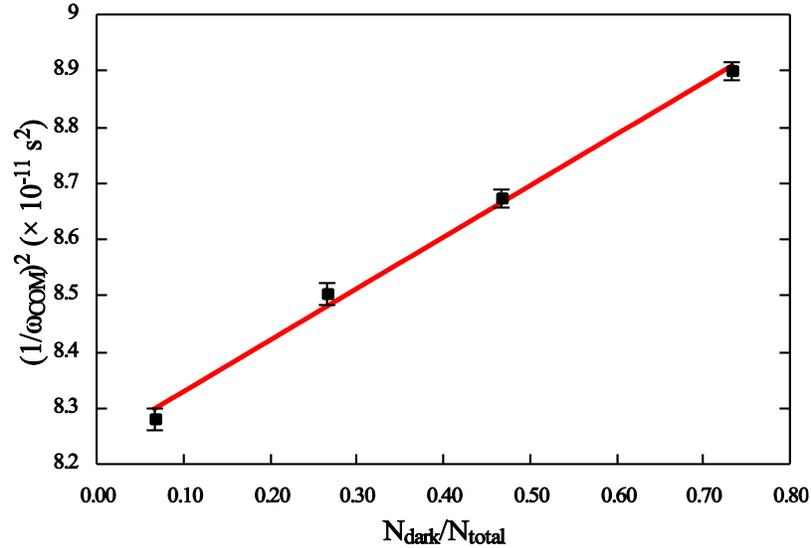

**Figure 6.** COM frequency versus chain composition. The linear fit expected by Eq. 8 from the measured frequency response of a bi-crystal yields a mass of 15.4 ± 0.8 amu for X in $BaX^+$, identifying the molecular ion, with high confidence, as $BaO^+$.

This dark-species identification method using low drive voltages complements another which employs higher rf drive voltages, in which the change in total fluorescence is monitored as the drive frequency is swept [26]. Note that these methods to identify the dark species require that the shift in resonant frequency caused due to reordering of the chain [7, 24] be much smaller than the shift associated with a composition change; this condition is met in the current work. Also, chain composition analysis using either method would become more complicated if there were more than one dark species in the chain or if the trapped species had different charges.

## 4. Conclusion

We have discussed the method for spatial-imaging thermometry for ions in a linear Coulomb crystal via fluorescence imaging of their spatial spreads. Imaging of each ion yields an independent measure of the crystal temperature, for the case that all normal modes are in thermal equilibrium at the same temperature. We experimentally show that the method works well for each of the ions in small chains of three ions. However, the technique breaks down for larger chains. Further investigation is required to determine whether patch potentials and mixing of axial and radial modes are indeed responsible for this breakdown of this thermometry method.

This technique offers an alternative method to more traditional fluorescence line profile measurements [27], which makes assumptions about the thermal distribution and requires precise knowledge of the saturation parameter and sweeping of the Doppler cooling laser frequency, and thus its cooling efficiency. Although not demonstrated here, the spatial-imaging thermometry technique has potential to make more accurate temperature measurements of linear Coulomb crystals, as it has previously been demonstrated to achieve for single trapped ions [11]. In addition, possible extensions of the technique to the case where a single temperature is not assumed for all normal modes of the chain might be used to investigate local heating effects. For instance, dark ions forced by light pressure to one end of the chain [4] would be expected to lead to non-uniform heating, which could potentially be characterized.

We also have demonstrated, using similar technology, a precise and easily implemented non-destructive technique for mass determination of dark ions in a two- species chain. Our approach to thermometry and mass spectrometry is applicable in general to particles undergoing Brownian motion that are coupled to each other and trapped in a harmonic oscillator potential. Our measurements using imaged spatial spreads open up possibilities for *in-situ* temperature measurement along with mass spectrometry in two-species chains.

## Acknowledgements

This work was supported by NSF Grant Nos. PHY-1404455 and DMR-1121262.